\documentclass[12pt]{iopart}
\usepackage{iopams} 
\usepackage{cite}
\usepackage{bm}% bold math
\usepackage{braket}
\usepackage{siunitx}
\usepackage{hyperref}
\usepackage{graphicx}
\usepackage{float}
\usepackage{xcolor}

\begin{document}

\title[Classical field simulation of vortex lattice melting in a 2D fast rotating Bose gas]{Classical field simulation of vortex lattice melting in a two-dimensional fast rotating Bose gas}

\author{S\'alvio J. Bereta$^{1,2}$, Lucas Madeira$^{1,3,4}$, Mônica A. Caracanhas$^1$, Hélène Perrin$^2$ and Romain Dubessy$^5$}

\address{$^1$ Instituto de F{\'\i}sica de S\~ao Carlos, Universidade de S\~ao Paulo, S\~ao Paulo, Brazil}
\address{$^2$ Université Sorbonne Paris Nord, Laboratoire de Physique des Lasers, CNRS UMR 7538, 99 av. J.-B. Clément, F-93430 Villetaneuse, France}
\address{$^3$ INFN-TIFPA Trento Institute of Fundamental Physics and Applications, Via Sommarive 14, I-38123 Trento, Italy}
\address{$^4$ European Centre for Theoretical Studies in Nuclear Physics and Related Areas (FBK-ECT*), Strada delle Tabarelle 286, Trento, Italy}
\address{$^5$ Aix-Marseille University, CNRS, PIIM, 13397, Marseille, France}
\ead{romain.dubessy@univ-amu.fr}
%\vspace{10pt}
%\begin{indented}
%\item[]August 2017
%\end{indented}

\begin{abstract}
We present a classical field simulation study of the thermal melting of a two-dimensional vortex lattice in a rotating Bose gas, focusing on the role of finite-size effects on the melting temperature. This work constitutes a numerical continuation of the recent experimental investigation reported in [Physical Review Letters \textbf{133}, 143401 (2024)], which addressed the thermal melting of a vortex lattice in a quasi-two-dimensional Bose gas. Using the stochastic projected Gross-Pitaevskii equation in a harmonic plus quartic trap, we simulate the finite-temperature equilibrium state and extract vortex configurations from density snapshots. Clear signatures of the two-step Kosterlitz--Thouless--Halperin--Nelson--Young melting scenario are identified. Our simulations enable a detailed characterization of the crystalline, hexatic, and liquid phases through correlation functions quantifying the translational and orientational order and through defect statistics. Finite-size effects are shown to play a crucial role at lower rotation frequencies, affecting the proliferation of lattice defects.
\end{abstract}

%
% Uncomment for keywords
%\vspace{2pc}
%\noindent{\it Keywords}: XXXXXX, YYYYYYYY, ZZZZZZZZZ
%
% Uncomment for Submitted to journal title message
\submitto{\NJP}
%
% Uncomment if a separate title page is required
%\maketitle
% 
% For two-column output uncomment the next line and choose [10pt] rather than [12pt] in the \documentclass declaration
%\ioptwocol
%

\section{Introduction}

The thermal melting of two-dimensional crystals is fundamentally different from the three-dimensional case and is commonly described within the Kosterlitz--Thouless--Halperin--Nelson--Young (KTHNY) framework \cite{Kosterlitz1972,Kosterlitz1973,Halperin1978,Nelson1979,Young1979}. In this scenario, the loss of crystalline order proceeds via two successive transitions at temperatures $T_{s/h} < T_{h/l}$, separating the solid, hexatic, and liquid phases. Below $T_{s/h}$, the equilibrium state is a triangular lattice exhibiting quasi-long-range translational order and long-range orientational order, with each lattice site having exactly six neighbors. When the temperature $T$ approaches $T_{s/h}$ from below, defects in the lattice appear in the form of bound pairs of dislocations, each dislocation being two adjacent sites with five and seven neighbors. As the temperature exceeds $T_{s/h}$, thermal fluctuations induce the unbinding of dislocation pairs. The resulting proliferation of free dislocations destroys translational order while preserving quasi-long-range orientational order, defining the hexatic phase. Above the second threshold $T_{h/l}$, another pair breaking mechanism occurs, isolated fivefold and sevenfold defects become unbound, leading to the loss of orientational order and the emergence of an isotropic liquid phase.

This two-dimensional melting scenario has been investigated experimentally in a wide variety of systems: colloidal solutions~\cite{Murray1987,Tang1989,Kusner1994,Marcus1996,Zahn1999,Han2008,Gasser2010,Kelleher2017}, air-fluidized dust or spheres~\cite{Petrov2015,Sun2016}, quantum vortex lattices in thin superconductors~\cite{Guillamon2009,Guillamon2014,Zehetmayer2015,Roy2019}, lattice of skyrmions~\cite{Huang2020}, and recently in a vortex lattice of a fast rotating superfluid~\cite{Sharma2024}. In parallel, intensive numerical simulations have been performed to test the melting scenario~\cite{Iaconis2010,Bernard2011,Wierschem2011,Kapfer2015,Li2019}, confirming the two-step melting process.

In this context, fast rotating superfluids are particularly interesting as they exhibit large vortex lattices, as observed in superfluid helium~\cite{Yarmchuk1979} and in dilute Bose-Einstein condensates~\cite{Abo-Shaeer2001,Coddington2003,Bretin2004}, within an exceptionally clean and highly controllable environment. For example, the trapping potential can be tailored to realize a quasi two-dimensional geometry, the temperature can be controlled by evaporative cooling, and the rotation frequency can be set with high precision. Furthermore, weakly interacting Bose gases can be described very accurately at low temperatures using classical field models that extend the mean-field zero-temperature Gross-Pitaevskii equation to finite temperature.

In this work, we use a classical field model, the stochastic projected Gross-Pitaevskii equation (SPGPE)~\cite{Gardiner2002,Gardiner2003}, to simulate the thermal equilibrium state of vortex lattices in a low-temperature, fast rotating Bose gas and study the vortex lattice melting scenario. Classical field models have primarily been used to study vortex-lattice formation in rotating Bose gases~\cite{Tsubota2002}, either by cooling a rotating thermal cloud~\cite{Penckwitt2002,Bradley2008} or by stirring a low-temperature condensate~\cite{Lobo2004}. This study is motivated by the fact that the experiment reported in Ref.~\cite{Sharma2024} observed a melting temperature significantly lower than an upper bound predicted by the KTHNY theory~\cite{Gifford2004}. Here, we aim to clarify the roles of finite-size effects, included in our model, in this discrepancy.

The paper is organized as follows. Section~\ref{sec:model} describes the SPGPE framework and numerical implementation. In Section~\ref{sec:results} we extract the melting temperatures and present the phase diagram. In Section~\ref{sec:discussion}, we compare these results with the analytical bounds of Ref.~\cite{Gifford2004}, and discuss finite-size and experimental effects. Section~\ref{sec:conclusion} provides concluding remarks and perspectives.

\section{Model and Numerical Methods}
\label{sec:model}

\subsection{Stochastic projected Gross--Pitaevskii equation}

To describe a finite-temperature Bose gas, we employ a classical-field simulation based on the stochastic projected Gross-Pitaevskii equation (SPGPE) in a harmonic plus quartic trap, within the simple-growth approximation~\cite{Bradley2008}. We expand the field
\begin{equation}
\psi_{\mathcal{C}}(\bm{r},t)=\sum_{n\in\mathcal{C}}c_n(t)\phi_n(\bm{r})   
\end{equation}
onto the single particle orbitals $\phi_n(\bm{r})$, where the set $\mathcal{C}=\{n~|~E_n<E_{\rm cut}\}$ contains the low-energy modes of the system up to the cutoff $E_{\rm cut}$.

The single particle orbitals $\phi_n(\bm{r})$ are obtained for the trapping potential
\begin{equation}
V(\bm{r})=\frac{M}{2}\left[\omega_r^2r^2\left(1+\kappa\frac{r^2}{a_r^2}\right)\right]+\frac{M}{2}\omega_z^2z^2,
\label{eqn:Vtrap}
\end{equation}
where $\omega_z$ is the oscillation frequency along the strongly confining vertical direction, assumed harmonic, $\omega_r$ is the oscillation frequency at the harmonic approximation in the radial direction, $a_r=\sqrt{\hbar/(M\omega_r)}$ and $\kappa$ is a small dimensionless parameter characterizing the quartic correction. We use a mixed Laguerre-Gauss and Hermite-Gauss basis~\cite{Bradley2008} adapted to account for the quartic term, see~\ref{app:SPGPE}. The spectrum also includes a term $-\Omega L_z$ which results from the change to the rotating frame, where $L_z$ is the projection along the symmetry axis $z$ of the angular momentum. 

The time evolution of the mode amplitudes $c_n(t)$ obeys the coupled non-linear equations
\begin{equation}
i\hbar\dot{c}_n(t)=(1-i\gamma)\left[(E_n-\mu)c_n(t)+g\int d\bm{r}\,\phi_n^*(\bm{r})|\psi_{\mathcal{C}}(\bm{r},t)|^2\psi_{\mathcal{C}}(\bm{r},t)\right]+\eta_n(t),
\label{eqn:SPGPE}
\end{equation}
where $g=4\pi\hbar^2a_s/M$ with $a_s$ the $s$-wave scattering length, $\gamma$ is a dimensionless damping coefficient, and $\mu$ is the chemical potential. The stochastic fields $\eta_n$ are Gaussian white-noise terms fixing the temperature through a fluctuation-dissipation relation:
\[
\braket{\eta_n^*(t)\eta_m(t^\prime)}=2\gamma\hbar k_BT\delta_{n,m}\delta(t-t^\prime),
\]
where $\delta_{n,m}$ is the Kronecker delta symbol and $\delta(t-t^\prime)$ is the Dirac delta distribution. In the right-hand side of Eq.~\eref{eqn:SPGPE} the overlap integral $\int d\bm{r}\,\phi_n^*(\bm{r})|\psi_{\mathcal{C}}(\bm{r},t)|^2\psi_{\mathcal{C}}(\bm{r},t)$ is evaluated exactly using the appropriate quadrature rule~\cite{Bradley2008}. Equation~\eref{eqn:SPGPE} is integrated using a second-order fixed-step stochastic integrator, with the white noise generated by a pseudorandom number generator.

\begin{figure}[ht]
    \centering
    \includegraphics[width=8.5cm]{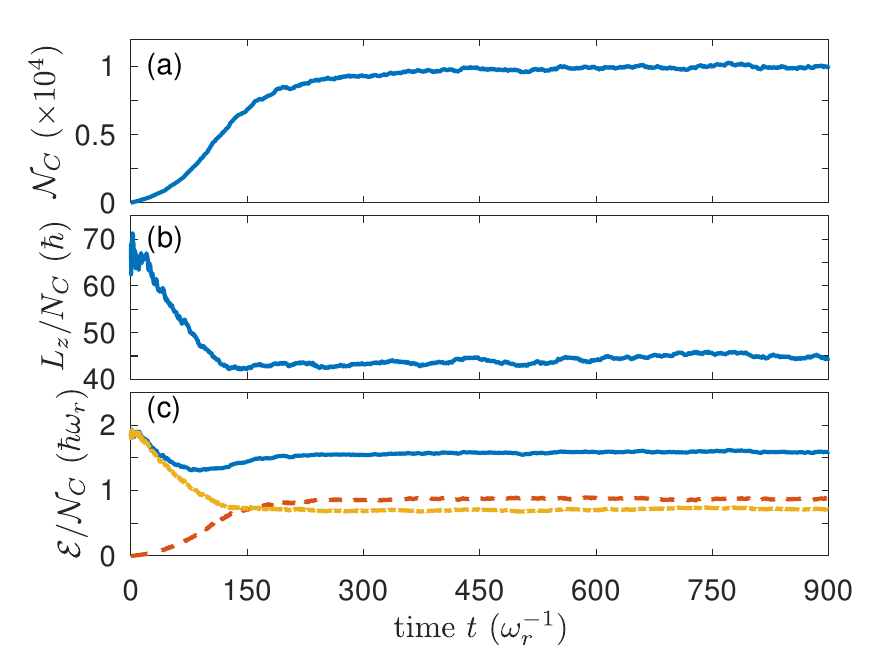}
    \caption{Convergence of the SPGPE simulation. Time evolution of (a) the coherent atom number $\mathcal{N}_C(t)$, (b) the angular momentum per particle $\mathcal{L}_z(t)/\mathcal{N}_C(t)$, and (c) the energy per particle $\mathcal{E}(t)/\mathcal{N}_C(t)$ for  $\Omega=0.99\omega_r$, $\mu=2.38\hbar\omega_r$, and $k_BT=3\hbar\omega_r$. In panel (c), the dashed red and dash-dotted yellow curves show the interaction and single-particle energy contributions, respectively.
    The steady-state coherent atom number is $\mathcal{N}_{\mathcal{C}}=9913\pm130$, with an estimated incoherent fraction $\mathcal{N}_{\mathcal{I}}\simeq356$, see text for details.}
    \label{fig:1}
\end{figure}

We solve Eq.~\eref{eqn:SPGPE} for a given choice of $\mu$, $\Omega$ and $T$ which are the relevant physical parameters, starting from a vacuum state $c_n(t=0)=0$. The classical field grows from random fluctuations and, after a transient, reaches a quasi-steady state in which the macroscopic quantities
\begin{eqnarray}
\mathcal{N}_{\mathcal{C}}(t)&=&\int d\bm{r}\,|\psi_{\mathcal{C}}(\bm{r},t)|^2,\\
\mathcal{L}_z(t)&=&\int d\bm{r}\,\psi_{\mathcal{C}}^*(\bm{r},t)L_z\psi_{\mathcal{C}}(\bm{r},t),\\
\mathcal{E}(t)&=&\sum_{n\in\mathcal{C}}E_n|c_n(t)|^2+\frac{g}{2}\int d\bm{r}\,|\psi_{\mathcal{C}}(\bm{r},t)|^4,
\end{eqnarray}
corresponding respectively to the atom number, angular momentum, and the total energy in the rotating frame, fluctuate around a steady-state value, as shown in Figure~\ref{fig:1}. Once this regime is reached, we assume ergodicity, meaning that the time evolution of the classical field samples the thermal equilibrium of the \emph{grand canonical} ensemble defined by $\{\mu,\Omega,T\}$, from which we study thermal equilibrium properties.

To satisfy the high occupation number assumption of the SPGPE model, we impose the energy cutoff by specifying a target occupation number $n_{\rm cut}$ at the cutoff, using the Bose-Einstein distribution for an ideal Bose gas: $E_{\rm cut}=\mu+k_BT\ln{(1+1/n_{\rm cut})}$.
In this work we take $n_{\rm cut}=2$.

\subsection{Simulation parameters}

We aim to model the experiment of Ref.~\cite{Sharma2024}, in which vortex-lattice melting was investigated for a $^{87}$Rb quantum gas with approximately constant atom number ($N\simeq 10^5\pm\SI{10}{\percent}$) confined in a quasi-2D harmonic plus quartic trap $(\omega_r,\omega_z,\kappa)=(2\pi\times\SI{34}{Hz},2\pi\times\SI{360}{Hz},\SI{1.5e-4}{})$. The rotation frequency was varied in the range $\Omega\in[0.7,1]\times\omega_r$ at constant temperature $T=\SI{18}{nK}$. For these parameters, the gas lies in the quasi-two dimensional regime, with a few excited axial harmonic oscillator levels populated, and always remains deep into the superfluid phase, $T/T_c<0.3$, where $T_c$ is the Berezinskii-Kosterlitz-Thouless critical transition temperature~\cite{Holzmann2008,Sharma2024}.

To perform the simulations we adopt several simplifying assumptions. First, we consider a system with $N=10^4$ atoms, therefore reaching the strictly two-dimensional SPGPE regime. This reduces the relevant energy and temperature scales compared with the experiment and significantly decreases the required computational basis size. Second, we study the transition as a function of the temperature for several rotation frequencies $\Omega$. For each pair $(\Omega,T)$, the chemical potential $\mu$ is tuned so that the steady-state coherent population $\mathcal{N}_{\mathcal{C}}$ is equal to the target atom number within a few percent.

In the simulations we use the radial trapping frequency $\omega_r$ as a reference scale for energies, times, frequencies, and $a_r$ as a length scale.
The large trap aspect ratio ($\omega_z/\omega_r\sim 10.6$) results in an effective two-dimensional interaction strength $g=\hbar^2\tilde{g}/M$, with $\tilde{g}=\sqrt{8\pi}a_s/a_z\simeq 0.0467$ and $a_z=\sqrt{\hbar/(M\omega_z)}$. Although the damping term $\gamma$ can be in principle derived from microscopic considerations, equilibrium properties are insensitive to its precise value provided $\gamma \ll 1$. We use $\gamma=0.01$ and we have tested that the results do not depend on this choice. The fixed time step $dt=0.01\,\omega_r^{-1}$ used in the integrator is chosen sufficiently small to accurately resolve the dynamics of the highest-energy modes. We always ensure that $dt E_{\rm cut}/\hbar\ll2\pi$.

Table~\ref{tab:1} reports the values we used for our simulations, the maximum temperature probed for each rotation frequency and the critical temperature in the harmonic plus quartic trap of Eq.~\eref{eqn:Vtrap}, in the rotating frame. The chemical potential tends to decrease weakly with the temperature, as we are keeping a fixed atom number, and we report here its mean value and variation over the range of temperatures simulated, from $T=0$ to $T_{\rm max}$.

\begin{table}[ht]
\centering
\caption{Values of $\Omega$, $\mu$, $E_{\rm cut}$ and the critical temperature $k_BT_c^{ho+q}$ in units of $\omega_r$ and $\hbar\omega_r$ respectively. Here we report only the energy cutoff value for the highest temperature $T_{\rm max}$.}
\label{tab:1}
\begin{tabular}{ccccc}%r}
\hline\hline
$\Omega/\omega_r$ & $\mu/\hbar\omega_r$ & $E_{\rm cut}/\hbar\omega_r$ & $k_BT_{\rm max}/\hbar\omega_r$ & $k_BT_{c}^{ho+q}/\hbar\omega_r$\\
\hline
0.95 & 4.21(2) & 6.6617 & 6.1 & 30.7 \\
0.96 & 3.85(4) & 6.2756 & 6.1 &  28.8  \\
0.97 & 3.44(2) & 5.5590 & 5.3  & 26.8 \\
0.98 & 2.97(1) & 4.3246 & 3.4  &  24.5\\
0.99 & 2.39(1) & 3.5964 & 3.0  &  21.8\\
1.00 & 1.7391(2) & 2.6709 & 2.3  & 17.8\\
\hline\hline
\end{tabular}
\end{table}

\subsection{Observables and analysis}

Simulations are initialized from a vacuum state and evolved for a total time $t_{\rm max}=900\,\omega_r^{-1}$, allowing the cloud to reach a steady state. We then extract 12 samples of the classical field at intervals of $10\,\omega_r^{-1}$, starting from $t=790\,\omega_r^{-1}$. Physical observables characterizing vortex lattice melting are computed from this sample set at each temperature, assuming ergodicity. For each parameter set $(\Omega,T,\mu)$, the procedure is repeated 10 times to construct an ensemble average over independent realizations and to estimate statistical uncertainties. 

\begin{figure}[ht]
    \centering
    \includegraphics[width=12cm]{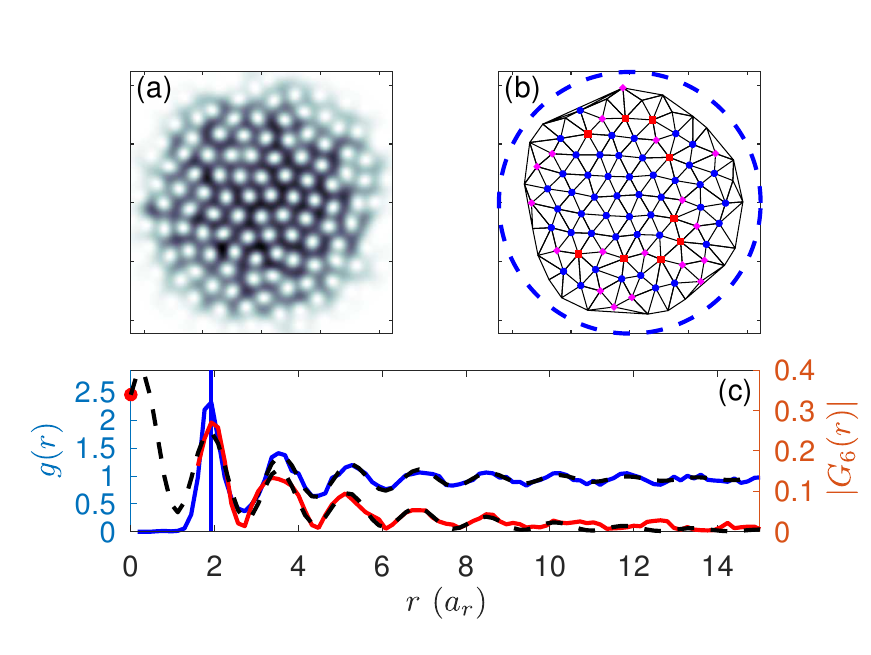}
    \caption{Example of finite-temperature vortex lattice. (a) Density profile. (b) Delaunay triangulation; the dashed blue circle indicates the Thomas-Fermi radius. (c) Pair correlation function $g(r)$ and orientational correlation function $G_6(r)$, averaged over $12\times10$ samples. The solid blue vertical line denotes the expected nearest neighbor spacing $a_v=\sqrt{2/\sqrt{3}n_v}$, where $n_v=M\Omega/\pi\hbar$ is the vortex density. The red dot indicates $|G_6(r=0)|$. Parameters: $\Omega=0.99\,\omega_r$, $\mu=2.38\,\hbar\omega_r$, and $k_BT=2.5\,\hbar\omega_r$. The dashed curves overlaid with the data are fits by damped cosine functions to extract the correlation lengths, see text for details.
    \label{fig:2}
    }
    \end{figure}

The main output of the simulation is snapshots of the density, see Fig.~\ref{fig:2}(a), and phase of the classical field. Vortex positions in each snapshot are identified from the local minima of the density. Because of the low signal-to-noise ratio near the cloud edge, the analysis is restricted to vortices inside a disk of radius $R=0.9\times R_{\rm TF}$, where
\begin{equation}
R_{\rm TF}=a_r\sqrt{\frac{\Omega^2/\omega_r^2-1+\sqrt{(\Omega^2/\omega_r^2-1)^2+8\kappa\mu/\hbar\omega_r}}{2\kappa}}
\label{eqn:RTF}
\end{equation}
is the Thomas-Fermi radius of a harmonic plus quartic trap obtained from a zero-temperature model~\cite{Cozzini2005,Cozzini2006}. Once the vortex positions are determined, a Delaunay triangulation is used to find nearest neighbors and reconstruct the vortex lattice, as shown in Figure~\ref{fig:2}(b).

To characterize the lattice state, we study several quantities. First, we compute the pair correlation function $g(r)$~\cite{Sharma2024}, which probes the translational order of the lattice. In the crystalline phase, $g(r)$ exhibits pronounced peaks at successive coordination shells (corresponding to first neighbor, second neighbor, ...) that progressively broaden and lose contrast as the temperature is increased, becoming essentially flat in the liquid phase. We fit $g(r)$ with an exponentially damped cosine model~\cite{Sharma2024} to extract the pair correlation length $\ell_P$.

Next, we compute the local orientational order parameter
\[
\psi_6(\bm{r}_k)=\frac{1}{N_k}\sum_{j=1}^{N_k}e^{6i\theta_{kj}},
\]
where the sum runs over the $N_k$ nearest neighbors of vortex $k$. The orientational correlation function is defined as
\[
G_6(r)=\braket{\psi_6^*(\bm{r}_k)\psi_6(\bm{r}_p)}_{|\bm{r}_k-\bm{r}_p|\sim r}.
\]
According to the KTHNY theory, the $G_6(r)$ function should exhibit a change of behavior at the hexatic to liquid transition, from an algebraic to an exponential decay~\cite{Kosterlitz1973}. We fit the $G_6(r)$ function with an exponential decay model to extract a correlation length $\ell_G$ of the orientational order. A typical example of the behavior of $g(r)$ and $G_6(r)$ is shown in Figure~\ref{fig:2}(c).

\begin{figure}[ht]
    \centering
    \includegraphics[width=12cm]{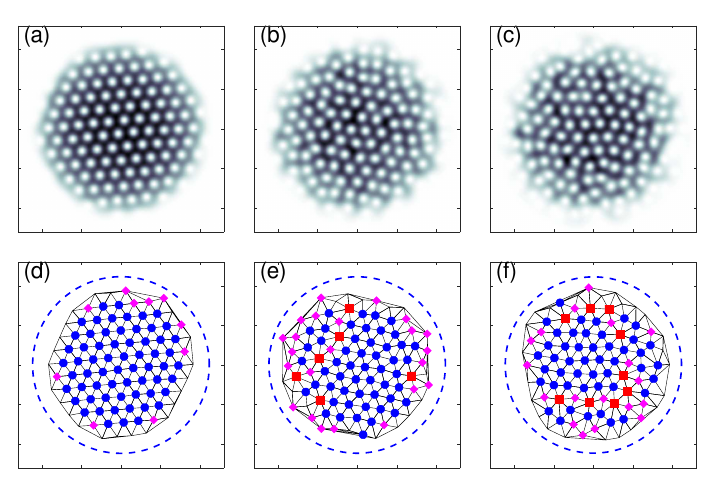}
    \caption{(a-c) Examples of density profiles and (d-f) corresponding vortex lattices at $\Omega=0.99\,\omega_r^{-1}$ for increasing temperatures $k_BT/\hbar\omega_r=\{0.1,1.4,2.5\}$ (left to right), in thermal equilibrium. The dashed blue circle indicates the Thomas-Fermi radius. In the vortex lattices, each site is labelled by its number of neighbors: blue disks for 6, red squares for 7 and pink diamonds for 5. The sequence illustrates the progression from crystalline to hexatic and liquid regimes (see Fig.~\ref{fig:4}).}
    \label{fig:3}
\end{figure}

Finally, we also quantify lattice order independently via the number of neighbors of each vortex. As shown in Figure~\ref{fig:3}, low-temperature lattices contain only sixfold coordinated sites in the bulk, while the number of defects (five- and sevenfold sites) increases with temperature. Because of the finite system size, the circular boundary of the condensate frustrates the vortex lattice and generates a high defect density close to the Thomas-Fermi radius. To remove this contribution, the defect analysis is restricted to vortices within a disk of radius $0.7\times R_{\rm TF}$.

\section{Results}
\label{sec:results}

\subsection{Melting transition temperatures}

\begin{figure}[ht]
\centering
\includegraphics[width=8.5cm]{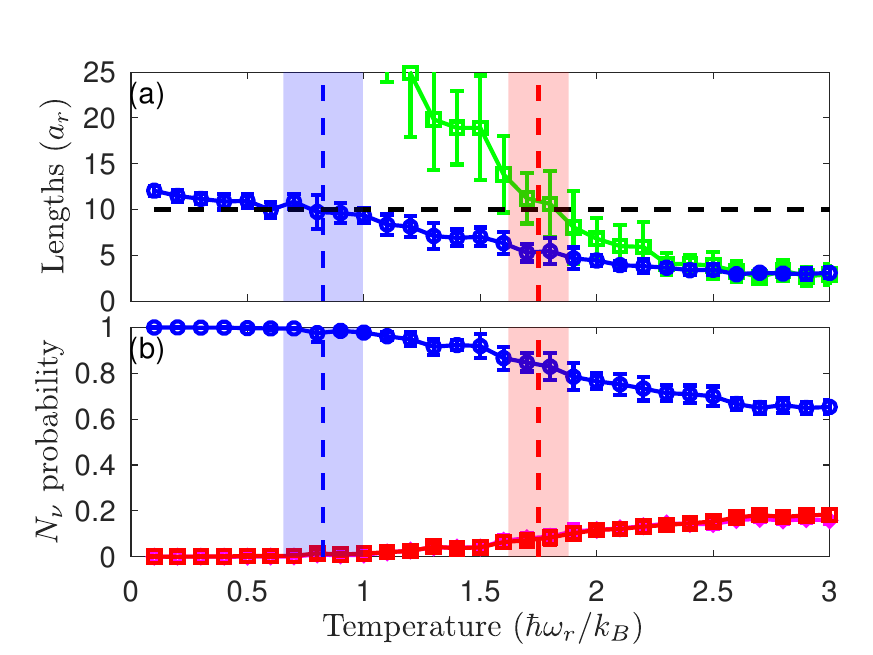}
\caption{(a) Correlation lengths $\ell_P$ (blue circles) and $\ell_G$ (green squares) as a function of temperature for $\Omega=0.99\,\omega_r$, computed within a disk of radius $R=0.9\times R_{\rm TF}$. The horizontal black dashed line indicates $R$. (b) Probability of having sites with 6 (blue circles), 5 (pink diamonds) and  7 (red squares) neighbors within a disk of radius $0.7\times R_{\rm TF}$.  In both panels, the vertical dashed lines mark the estimated transition temperatures, and the shaded areas the uncertainties, see text for details. The error bars correspond to statistical uncertainties.}
\label{fig:4}
\end{figure}

Figure~\ref{fig:4} shows the analysis of the vortex lattice order for $\Omega=0.99\,\omega_r$ as a function of the temperature $T$. The pair and orientational correlation lengths, $\ell_P$ and $\ell_G$, tend to decrease with increasing temperature. According to the KTHNY theory of melting, the translational order is lost first, as seen from the decrease of $\ell_P$, while the number of defects tends to increase. At higher temperature, $\ell_G$ drops sharply and the vortex lattice reaches a completely disordered state with a large number of bulk defects. Five- and sevenfold-coordinated sites appear in pairs, as expected for a dislocation-disclination-mediated melting mechanism.

To define the melting temperatures $T_{s/h}$ and $T_{h/l}$, we introduce combined criteria based on correlation lengths and defect proliferation. The two criteria are combined to mitigate ambiguities due to finite-size effects and length uncertainties, ensuring a more robust estimate. In particular, we use the pair and orientational correlation lengths, $\ell_P$ and $\ell_G$, together with the fraction of sevenfold coordinated sites ($f_7$). The length $\ell_P$ primarily probes translational order and is therefore associated with the crystal–hexatic transition, whereas $\ell_G$ probes orientational order and characterizes the hexatic–liquid transition. The critical temperature $T_{s/h}$ is obtained by taking the average of the datasets satisfying
\begin{equation}
T_{s/h}:  \quad R \in [\ell_P - \delta \ell_P, \ell_P + \delta \ell_P] \quad \text{and} \quad f_7 + \delta f_7 < 5\%, \label{eq:Tsh}
\label{eqn:Tsh}
\end{equation}
and, analogously, $T_{h/l}$ from
\begin{equation}
T_{h/l}:  \quad R \in [\ell_G - \delta \ell_G, \ell_G + \delta \ell_G] \quad \text{and}  \quad f_7 - \delta f_7 > 5\%.\label{eq:Thl}
\label{eqn:Thl}
\end{equation}
Here $R=0.9\times R_{\rm TF}$ denotes the effective system size, and $\delta \ell_{P,G}$ and $\delta f_7$ represent statistical uncertainties. We focus on the probability of finding a site with 7 neighbors because, at lower rotation frequencies and for smaller lattices, finite-size effects tend to artificially introduce sites with 5 neighbors on the boundary, even in the crystalline phase, see~\ref{app:set}. The uncertainties on $T_{s/h}$ and $T_{h/l}$ were estimated by computing the standard deviations of the datasets defined by Eqs.~\eref{eqn:Tsh} and \eref{eqn:Thl}, respectively. 

\begin{figure}
    \centering
    \includegraphics[width=8.5cm]{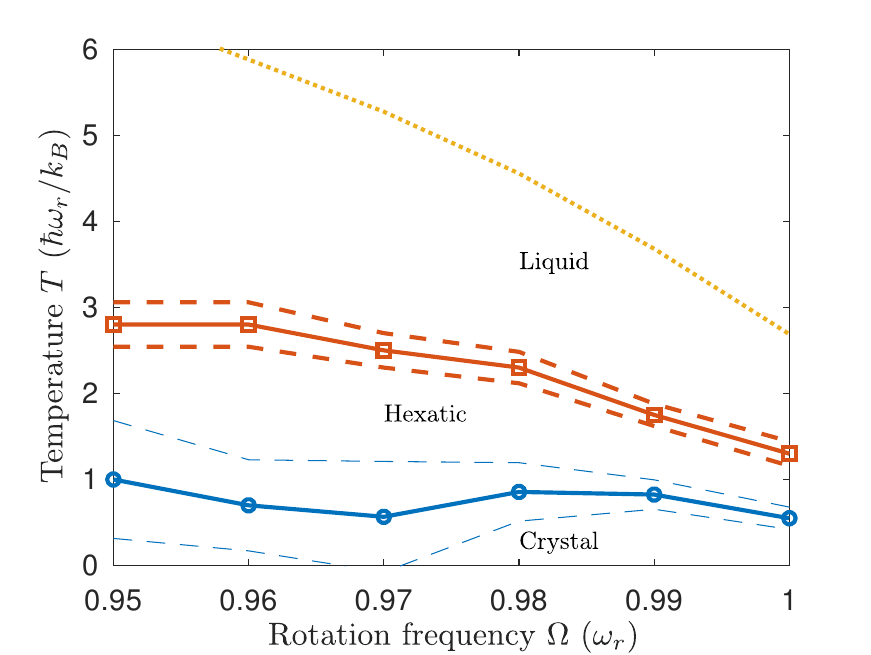}
    \caption{Vortex lattice equilibrium phases as a function of the rotation frequency and temperature for $N=10^4$. Blue circles (orange squares) indicate the estimated crystal-hexatic $T_{s/h}$ (hexatic-liquid $T_{h/l}$) transition temperatures obtained from the simulations. The solid lines connect the points as a guide to the eye. The yellow dotted curve is an upper bound for the melting temperature, see Eq.~\eref{eq:upperbound}. Dashed curves indicate the estimated uncertainty on the transition temperatures (see text for details).}
    \label{fig:5}
\end{figure}

We have repeated the same protocol for several rotation frequencies, and extracted the corresponding transition temperatures $T_{s/h}$ and $T_{h/l}$, as shown in Fig.~\ref{fig:5}. Over the explored range, both $T_{s/h}$ and $T_{h/l}$ decrease with increasing rotation frequency, as expected, with a stronger dependence for $T_{h/l}$. The crystal-hexatic transition is more difficult to determine precisely, as reflected by the larger uncertainties on $T_{s/h}$. We attribute this to the finite size of the system, which smoothens the crossovers between different phases. Nevertheless, we can clearly identify the three different phases in our simulations.

In Fig.~\ref{fig:5}, the simulated rotation frequency range is limited to $\Omega/\omega_r\in[0.95,1]$, mainly due to finite-size effects. Requiring that the zero-temperature equilibrium state forms a well-defined vortex lattice with at least ten sites across the cloud diameter sets a lower bound on the rotation frequency. This bound can be estimated from the equilibrium vortex density $n_v=M\Omega/\pi\hbar$ and the Thomas-Fermi radius, Eq.~\eref{eqn:RTF}, for a given atom number (or chemical potential). For our simulation parameters ($N=10^4$), this yields $\Omega/\omega_r\geq0.95$. Below this value, thermal fluctuations still affect the positions of the vortices, but the limited lattice size makes it difficult and less meaningful to distinguish phases with long-range correlations. Reducing this bound would require decreasing $\kappa$ or increasing the atom number: for example, with $N=10^5$ the relevant range becomes $\Omega/\omega_r\geq0.66$, consistent with the regime explored experimentally in Ref.~\cite{Sharma2024}.

\subsection{Theoretical bounds for the melting temperature}

The simplest estimate of an upper bound for the melting temperature of a fast-rotating two-dimensional weakly interacting Bose gas is given in Ref.~\cite{Gifford2008}:
\begin{equation}
    k_BT_m = \frac{1}{8\sqrt{3}}\frac{\mu}{\tilde{g}}.
    \label{eq:upperbound}
\end{equation}
This bound is derived from the known value of the vortex lattice shear modulus in the incompressible regime ($\hbar\Omega\ll\mu$) and neglects the presence of an intermediate hexatic phase. Using the chemical potential $\mu(\Omega)$ obtained from the simulations, we plot the upper bound of Eq.~\eref{eq:upperbound} in Fig.~\ref{fig:5}. Over the explored rotation frequency range, the bound is approximately twice the observed melting temperature $T_{h/l}$ (hexatic-liquid transition). This is consistent with experimental findings~\cite{Sharma2024}, where the upper bound significantly overestimated the actual melting temperature.

We may also evaluate Eq.~\eref{eq:upperbound} directly from the model parameters. Since we are in the very low-temperature regime, the chemical potential can be approximated by the Thomas-Fermi prediction for a two-dimensional Bose gas in a harmonic plus quartic trap~\cite{Cozzini2006}, with an effective interaction strength $b_\Delta\tilde{g}$. This interaction is renormalized by the Abrikosov parameter $b_\Delta\simeq1.16$ to account for coarse graining over the vortex core size~\cite{Gifford2008}. We find that the chemical potential obtained from the simulations is in good agreement with the zero-temperature Thomas-Fermi estimate, thus supporting the validity of this approximation.

\section{Discussion}
\label{sec:discussion}

The first outcome of this work is that the SPGPE simulations of vortex lattice melting seem to follow closely the KTHNY scenario. As discussed above, finite-size effects associated with the small lattice sizes are significant. Nevertheless, we obtain clear evidence for a two-stage melting process as predicted by the KTHNY theory. To the best of our knowledge, this work is the first study of vortex lattice melting in a fast-rotating superfluid using a classical field model. From the shape of the phase boundaries in Fig.~\ref{fig:5}, it is apparent that resolving the two transitions by varying the rotation frequency at fixed temperature, as in Ref.~\cite{Sharma2024}, is extremely challenging. These results therefore suggest that improved experimental control enabling measurements as a function of temperature at fixed rotation frequency would contribute to a clearer characterization of the melting sequence.

Since vortex lattice melting occurs at very low temperatures, our simulations employ the SPGPE method in a parameter regime where its formal validity is uncertain. In the microscopic derivation of the SPGPE model, the high occupation number assumption of the classical field model usually requires $k_BT\gg\mu$, a condition not satisfied in our simulations. Although our results show that the SPGPE dynamics follow a KTHNY melting scenario, the question of its quantitative relevance for describing a real dilute, weakly interacting Bose gas in this regime remains open.

To further investigate this question, it would be interesting to perform a direct comparison with the experiment of Ref.~\cite{Sharma2024}. This would require simulations of larger systems with higher atom numbers in the quasi-two-dimensional regime, which are computationally more demanding. Our implementation of the SPGPE equations already accounts for the third dimension, that will be included in the computational basis if $E_{\rm cut}>\hbar\omega_z$, see \ref{app:SPGPE}. More generally, the SPGPE equation could be used to study the two- to three-dimensional crossover and track the evolution of the melting temperature with the condensate aspect ratio. We know from the upper bound estimate~\cite{Gifford2008} that the melting temperature should get closer to the critical temperature for Bose-Einstein condensation in a three dimensional trapping geometry. This would enable tests of the melting scenario in that regime and allow investigation of competing processes affecting vortex lines, such as Kelvin modes~\cite{Chevy2003,Rooney2011}. In principle, it should be possible to find a set of parameters in which the SPGPE formalism is fully appropriate for describing the finite-temperature equilibrium properties of the system.

As discussed above, the upper bound of Eq.~\eref{eq:upperbound} largely overestimates the observed melting temperature. We currently have no clear explanation for the origin of this discrepancy. We may question the validity of the incompressible lattice assumption; however, it has been shown~\cite{Cozzini2006b} that the vortex lattice shear modulus remains close to the incompressible value up to $\hbar\Omega\simeq\mu$. Alternatively, we may argue that finite-size effects play an important role, given the moderate lattice sizes in our simulations. A systematic study of this effect would require varying the system size at fixed coarse-grained density (or equivalently fixed chemical potential). This could be achieved, for example, by setting $\Omega=\omega_r$ and varying the quartic coefficient $\kappa$, since $R_{\rm TF}(\Omega=\omega_r)=a_r(2\mu/(\hbar\omega_r\kappa))^{1/4}$. However this remains numerically demanding, as doubling the system size requires reducing $\kappa$ by a factor of 16. Furthermore, the melting temperature is expected to decrease with increasing system size~\cite{Iaconis2010}, which would further enlarge the discrepancy with the upper bound prediction.

Finally, it would be interesting to fix the system size and vary the chemical potential. By reducing $\mu$, the system could be driven progressively into the lowest Landau level (LLL) regime within mean-field theory. This may help establish an improved upper bound for the melting temperature, since the special form of the LLL wave functions is convenient for analytical treatment.

\section{Conclusion}
\label{sec:conclusion}

In this work, we presented classical field simulations of the thermal equilibrium state of a fast-rotating two-dimensional Bose gas using the SPGPE formalism. The numerical results provide clear evidence of a two-step melting scenario in agreement with the KTHNY theory. The observed melting temperature is a factor of two lower than the upper bound reported in Ref.~\cite{Gifford2008}, in line with the trend observed experimentally~\cite{Sharma2024}. Further numerical simulations are necessary to clarify the origin of this discrepancy and possibly to establish an improved estimate of the melting temperature.

The SPGPE formalism is a convenient framework for testing the KTHNY scenario and studying, for example, the dimensional crossover from three to two dimensions, melting in the lowest Landau level regime, or the impact of finite-size effects. However, further studies are needed to assess its validity at such low temperatures, in particular through comparisons with other methods.

Finally, it would be interesting to study how temperature quenches across the freezing and melting transitions can be modeled within the SPGPE framework, and to test if the number of defects depends on the quench rate, as predicted by the Kibble-Zurek theory~\cite{Kibble1976,Zurek1985}, at least within the numerical model.

\ack
LPL is UMR 7538 of CNRS and Sorbonne Paris Nord University.
We acknowledge financial support from the ANR project VORTECS (Grant No. ANR-22-CE30-0011) and USP-COFECUB (Grant No. Uc Ph 177/19).
This work was supported by the São Paulo Research Foundation (FAPESP) under the grants 2013/07276-1, 2024/04637-8, and 2024/19338-6, and CAPES-PRINT (financial code 0001) No.88887.695330/2022-00. 
R.D. and H. P. acknowledge support from the French government under the France 2030 investment plan, as part of the Initiative d'Excellence d'Aix-Marseille Université -- AMIDEX AMX-22-CEI-069 and in the framework of PEPR project QUTISYM -- ANR-23-PETQ-0002.

\appendix

\section{Implementation of the SPGPE}
\label{app:SPGPE}
The key ingredients of the SPGPE method are the single-particle orbitals and spectrum. For a harmonic trap in a rotating frame, the orbitals can be constructed using Laguerre-Gauss and Hermite-Gauss basis~\cite{Bradley2008}:
\[
\phi_n(\bm{r})=\frac{e^{-z^2/2a_z^2}}{\sqrt{n_z!\sqrt{\pi}2^{n_z}}}\,H_{n_z}\!\!\left(\frac{z}{a_z}\right)\times 
e^{i\ell\theta}\sqrt{\frac{n_r!}{\pi(n_r+|\ell|)!}}\frac{r^{|\ell|}}{a_r^{|\ell|}}\,e^{-\frac{r^2}{2a_r^2}}L_{n_r}^{|\ell|}\!\left(\frac{r^2}{a_r^2}\right),
\]
where $H_{n_z}$ is a Hermite-Gauss polynomial and $L_n^{|\ell|}$ is a Laguerre-Gauss polynomial, parametrized by the three integers $n\equiv(n_r,\ell,n_z)\in\mathbb{N}\times\mathbb{Z}\times\mathbb{N}$.
The spectrum is the one of the harmonic oscillator:
\[
E_n^{ho}=\hbar\omega_r(2n_r+|\ell|)-\hbar\Omega\ell+\hbar\omega_z n_z,
\]
where we have removed the zero-point energy $\hbar\omega_r+\hbar\omega_z/2$.
The additional quartic term in Eq.~\eref{eqn:Vtrap} mixes states with different $n_r$ indices but same $(\ell,n_z)$. To lowest order, the single particle spectrum is modified as:
\[
E_n^{approx}=E_n^{ho}+\kappa\frac{\hbar\omega_r}{2}(\ell^2+(3+6n_r)\ell+2+6n_r(n_r+1)).
\]
For a particular choice of $(\Omega,\mu,T)$, we use this approximate energy spectrum to define which orbitals of the bare harmonic oscillator are below the energy cutoff, included in the set: $n_z\leq E_{\rm cut}/\hbar\omega_z$, $n_r\leq E_{\rm cut}/2\hbar\omega_r$, and
\[
 |\ell|-\frac{\Omega}{\omega_r}\ell+\frac{\kappa}{2}(\ell^2+3\ell+2)\leq \frac{E_ {\rm cut}}{\hbar\omega_r}.
\]
Because this expression for the spectrum is only approximate, we include in the initial basis more states, corresponding to higher $n_r$ and $\ell$ values (typically 10 extra states). We then evaluate exactly the quartic correction in this basis, by diagonalizing the single-particle Hamiltonian. In this way, we obtain an accurate description of the single particle spectrum $E_n$ corresponding to the potential of Eq.~\eref{eqn:Vtrap}, as well as the modified single particle orbitals. Finally, we keep for the computation only the states with an energy below the cutoff. Since these orbitals are linear combinations of the original basis function, the non-linear term in Eq.~\eref{eqn:SPGPE} can still be evaluated exactly using an appropriate quadrature. All these operations can be efficiently implemented using matrix operations~\cite{Bradley2008}.

We implement the SPGPE, Eq.~\eref{eqn:SPGPE}, using our own code, written in Octave programming language~\cite{Eaton2025}, using dimensionless units: $r\to\tilde{r}a_r$, $z\to\tilde{z}a_z$, $t\to\tilde{t}/\omega_r$, $\phi_n(\bm{r})\to\tilde{\phi}_n(\tilde{\bm{r}})/\sqrt{a_r^2a_z}$, $\psi_{\mathcal{C}}(\bm{r},t)\to\tilde{\psi}_{\mathcal{C}}(\tilde{\bm{r}},\tilde{t})/\sqrt{a_r^2a_z}$, $E_n\to\hbar\omega_r\tilde{E}_n$, $\mu\to\hbar\omega_r\tilde{\mu}$, and $\eta_n(t)\to\tilde{\eta}_n(\tilde{t})\hbar\omega_r$ such that Eq.~\eref{eqn:SPGPE} reads:
\begin{eqnarray*}
&&i\frac{dc_n(\tilde{t})}{d\tilde{t}}=(1-i\gamma)\left[(\tilde{E}_n-\tilde{\mu})c_n(\tilde{t})+\frac{4\pi a_s}{a_z}\int d\tilde{\bm{r}}\,\tilde{\phi}_n(\tilde{\bm{r}})^*|\tilde{\psi}_{\mathcal{C}}(\tilde{\bm{r}},\tilde{t})|^2\tilde{\psi}_{\mathcal{C}}(\tilde{\bm{r}},\tilde{t})\right]+\tilde{\eta}_n(\tilde{t}),\\
&&\braket{\tilde{\eta}_n(\tilde{t})^*\tilde{\eta}_m(\tilde{t}^\prime)}=2\gamma\frac{ k_BT}{\hbar\omega_r}\delta_{n,m}\delta(\tilde{t}-\tilde{t}^\prime).
\end{eqnarray*}
It is then natural to define the temperature in units of radial trap energy.

\section{Complete data set}
\label{app:set}

\begin{figure}
    \centering
    \includegraphics[width=7.5cm]{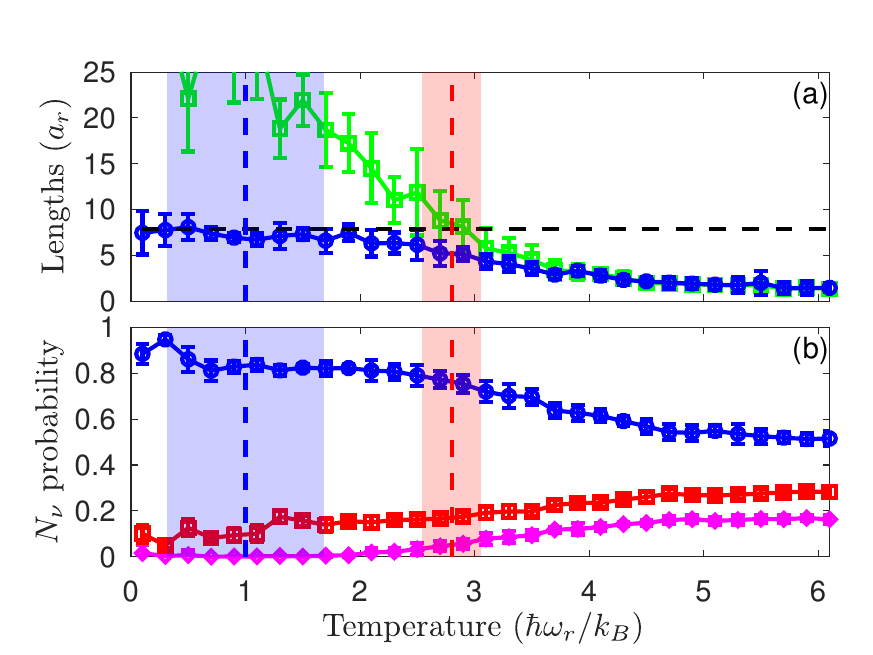}
    \includegraphics[width=7.5cm]{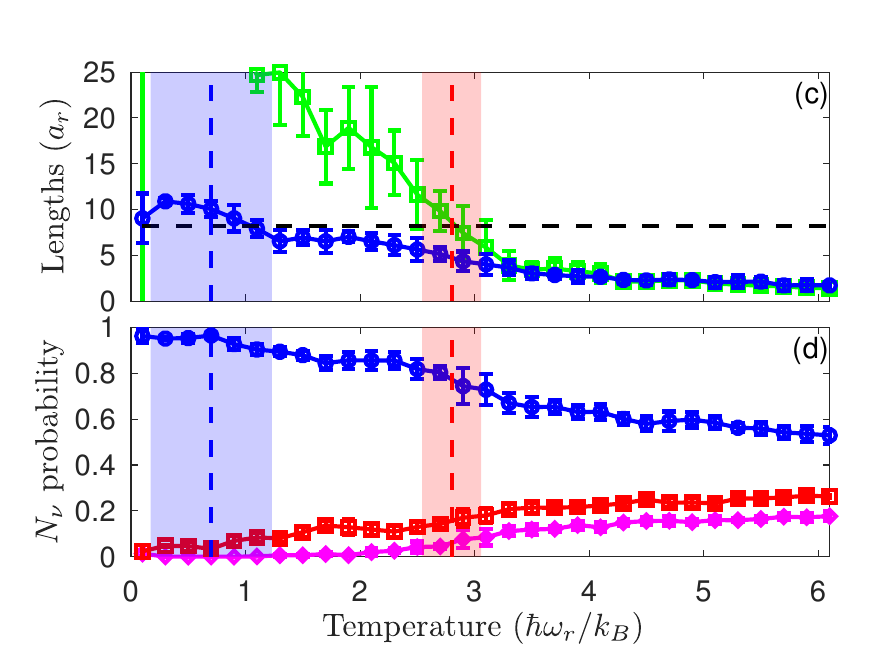}\\
    \includegraphics[width=7.5cm]{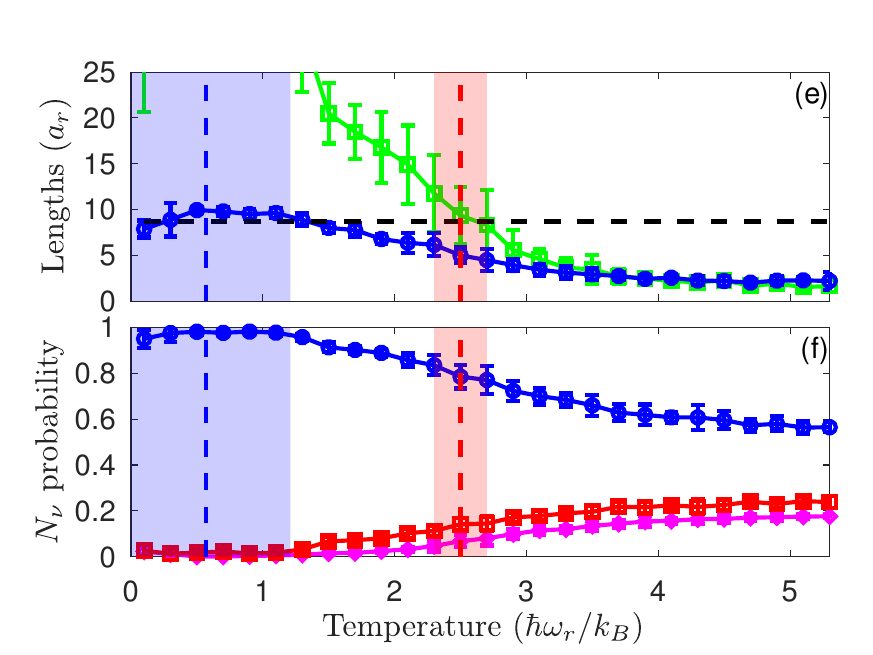}
    \includegraphics[width=7.5cm]{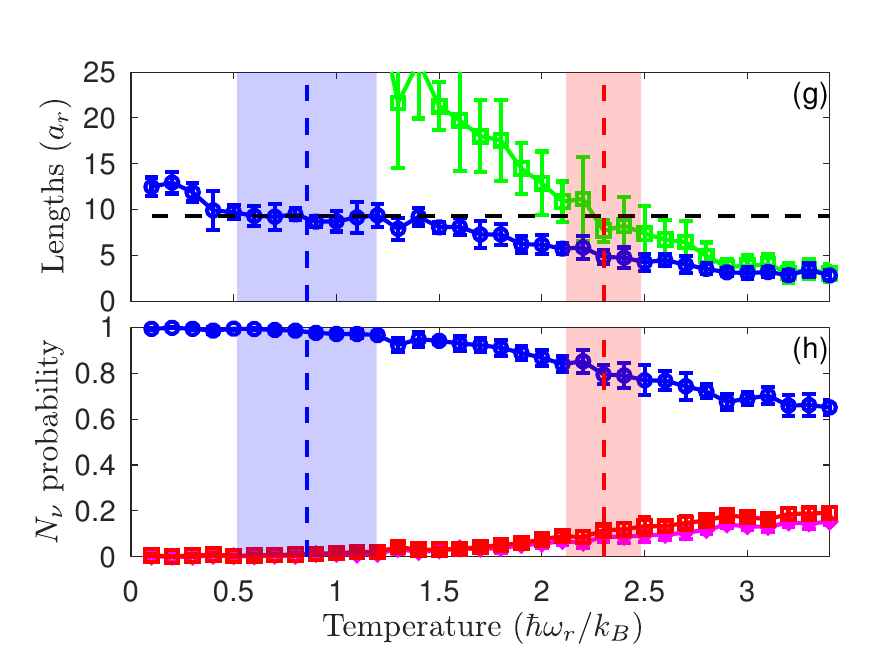}\\
    \includegraphics[width=7.5cm]{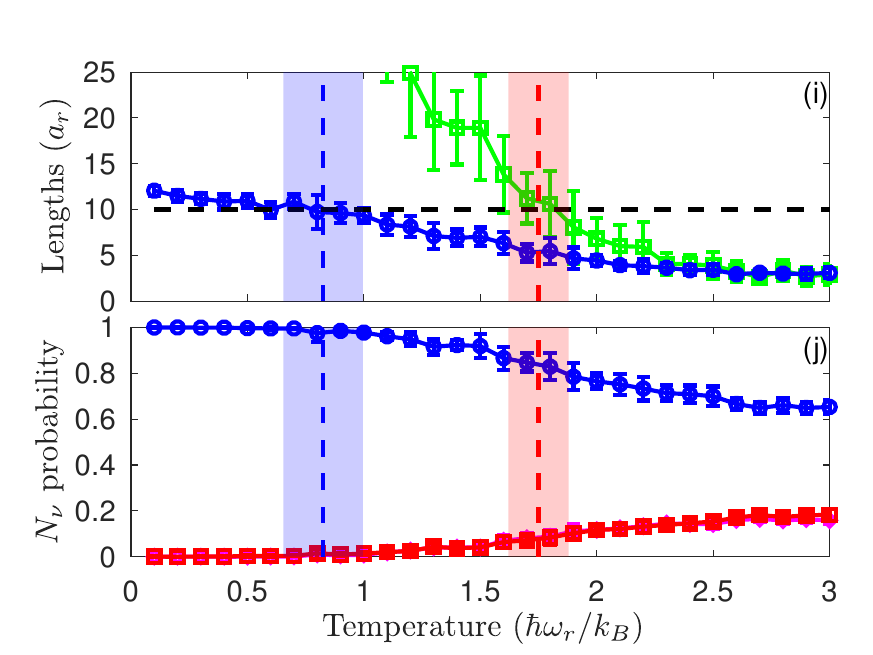}
    \includegraphics[width=7.5cm]{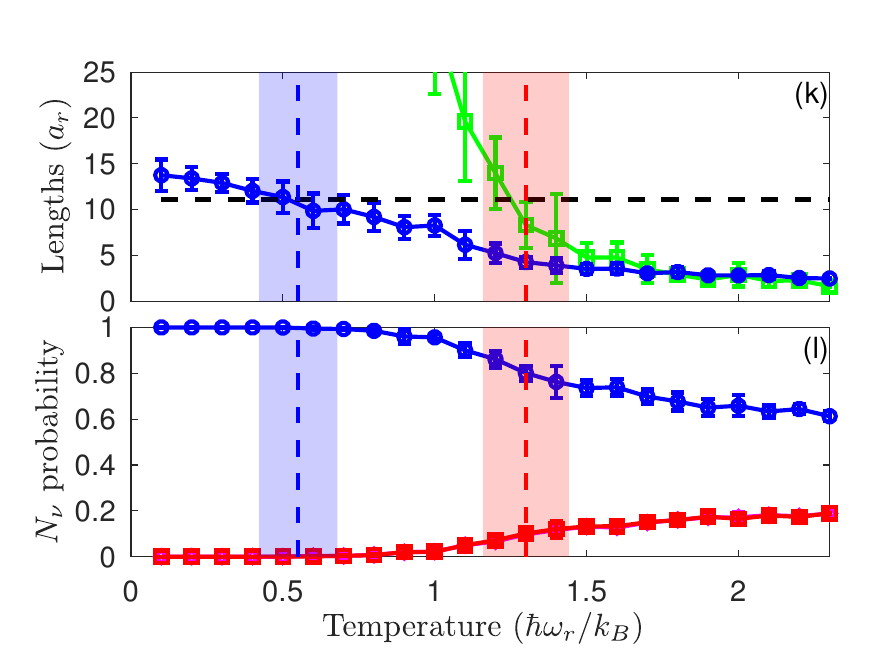}
    \caption{(a,c,e,g,i,k) Correlation lengths $\ell_P$ (blue circles) and $\ell_G$ (green squares) as a function of temperature for several rotation frequencies $\Omega$, computed within a disk of radius $R=0.9\times R_{\rm TF}$. The horizontal black dashed line indicates $R$. (b,d,f,h,j,l) Probability of having sites with 6 (blue circles), 5 (pink diamonds) and 7 (red squares) neighbors within a disk of radius $0.7\times R_{\rm TF}$. In all panels, vertical dashed lines mark the estimated transition temperatures and shaded areas their uncertainties; error bars denote statistical uncertainties. (a-b) $\Omega=0.95\,\omega_r$, (c-d) $\Omega=0.96\,\omega_r$, (e-f) $\Omega=0.97\,\omega_r$, (g-h) $\Omega=0.98\,\omega_r$, (i-j) $\Omega=0.99\,\omega_r$, (k-l) $\Omega=\omega_r$.}
    \label{fig:B1}
\end{figure}

Figure~\ref{fig:B1} shows the analysis of the thermal equilibrium states in the vicinity of the melting transition, for rotation frequencies $\Omega/\omega_r\in[0.95,1]$. The figures are obtained and analyzed with the same methods as for Fig.~\ref{fig:4} in the main text, see section \ref{sec:results}. For the sake of clarity and completeness, we included in this comparison the value $\Omega=0.99\,\omega_r$, such that Fig.~\ref{fig:4}(a-b) and Fig.~\ref{fig:B1} (i-j) are identical.

\section*{References}
\bibliographystyle{iopart-num}
\bibliography{biblio}% Produces the bibliography via BibTeX.

\end{document}